\documentstyle[11pt,aaspp4]{article}

\tighten

\slugcomment{Accepted for publication in Astrophysical Journal Letters}

\begin{document}

\title{The Rise Times and Bolometric Light Curve of SN 1994D:\\
       Constraints on Models of Type Ia Supernovae}

\author{William D. Vacca\altaffilmark{1}}
\affil{Institute for Astronomy, University of Hawaii, \\
       2680 Woodlawn Drive, Honolulu, HI 96822 \\
       e-mail: vacca@athena.ifa.hawaii.edu}
\altaffiltext{1}{Beatrice Watson Parrent Fellow}

\and
\author{Bruno Leibundgut}
\affil{European Southern Observatory, \\
       Karl-Schwarzschild-Strasse 2, D-85748, Garching, Germany\\
       e-mail:bleibundgut@eso.org}

\begin{abstract}
Using the published photometry and an empirical model of the temporal evolution
of the apparent magnitudes in the {\it UBVRI} passbands, we have constructed 
a continuous optical bolometric, or ``quasi-bolometric'', light curve of the 
well-observed Type Ia supernova SN 1994D. The optical bolometric light curve is 
found to have a maximum luminosity of about 
$8.8\times 10^{42} \times (\frac{D}{13.7 Mpc})^2$ erg/s, which is reached 
$\sim 18$ days after the explosion. In addition, the optical bolometric light curve 
exhibits an inflection, or ``shoulder'', about $25$ days after maximum. This inflection
corresponds to the secondary maximum observed in all filter light curves redder than $B$. 
The individual filter curves have rise times similar to that of the optical bolometric 
light curve; other Type Ia supernovae are found to have similar rise times.  
Our fits indicate that the peak bolometric luminosity and the maxima in the   
{\it B}, {\it V}, and {\it R} light curves all occur within a day of one another. 
These results can be used to place constraints on theoretical models of Type Ia events.
For example, all current theoretical models predict rise times to peak luminosity
which are significantly shorter than that estimated for SN 1994D.

\end{abstract}


\section{Introduction}

To compare theoretical models for the temporal evolution of the emitted flux 
from Type Ia supernovae (SNe Ia) with observations, it is necessary either to calculate 
the predicted light curves in the various observational passbands from the models or
to construct bolometric light curves from the data. Attempts have been made to calculate 
the filter light curves from models with elaborate hydrodynamics codes (e.g. H\"oflich
\& Khokhlov 1996; Eastman 1996), but a consensus on the correct treatment of the
radiation transfer has yet to be reached (Eastman 1996; Pinto 1996; H\"oflich 1996).
Accurate filter light curves are difficult to calculate in this manner, as they
require synthesization of the entire spectrum on a large number of wavelength points 
at every stage in the evolution of the explosion. On the other hand, the construction 
of truly bolometric light curves from observations requires numerous simultaneous 
measurements across the entire electromagnetic spectrum, including the satellite 
ultraviolet, the optical {\it UBVRI} passbands, the near-infrared {\it JHK} passbands, 
and longer wavelengths. No SN Ia event has ever been observed extensively in all of 
these bands; few have a substantial number of observations even in the optical passbands. 
Nevertheless, with a few assumptions regarding the amount of energy emitted at  
unobserved wavelengths, approximate bolometric light curves can be derived from 
observed photometric data. In this manner, preliminary bolometric light curves for 
a few well-observed SNe~Ia (e.g., SN 1992A) have been constructed by Suntzeff (1996). 
As an observationally-derived estimate of the total thermal energy radiated from a 
SN Ia, a bolometric light curve provides a crucial link between theory and observations  
by allowing model predictions to be directly compared with observational results.
The value of a bolometric light curve assembled from observations to constrain 
SN explosion models has been clearly demonstrated in the case of SN~1987A 
(Suntzeff et al. 1991; Bouchet, Danziger, \& Lucy 1991).

SN 1994D was a Type Ia event that occurred in NGC 4526, an S0 galaxy located 
about $5^{\circ}$ from M87 in the Virgo cluster.\footnote[2]{With a velocity of 
$\sim 533$ km/s (Binggeli et al. 1985), NGC 4526 is clearly a member of the Virgo 
cluster. Recent reports of the distance modulus for the Virgo cluster range from 
31.1 (Freedman et al. 1994, Saha et al. 1996) to 32.0 (Sandage et al. 1996). We will 
adopt the distance modulus for NGC 4526 of 30.68 (13.7 Mpc) favored by Richmond et al. 
(1995) and Patat et al. (1995) but will scale the luminosity accordingly.} Accurate and 
extensive observations of SN~1994D have been reported by Richmond et al. (1995), Patat 
et al. (1995), and Meikle et al. (1996). H\"oflich (1995) compared optical filter light 
curves predicted  from theoretical models with unpublished photometry from CTIO and the 
CfA. Although the models were able to reproduce the general characteristics of the filter 
light curves, the detailed shapes of the curves could not be matched. 

SN~1994D was an unusual SN~Ia in several respects; it displayed an unusually blue color 
near maximum, appeared to be overluminous at maximum compared to other SNe Ia, and deviated 
from the decline-luminosity relation observed for Type Ia events (Richmond et al. 1995, 
Patat et al. 1995). Near maximum, the spectrum of SN~1994D exhibited a strong P-Cygni 
line near 1.05$\mu$m which is possibly due to He~I ($\lambda_0$=1.083$\mu$m;
Meikle et al. 1996). The degree to which SN~1994D represents a truly peculiar SN~Ia, as 
opposed to a minor variant of a ``normal'' SNe Ia, is still to be determined, however. 
In particular, the derived luminosity is somewhat uncertain due to uncertainty in the 
distance to NGC 4526, and the spectral region observed by Meikle et al. (1996) has not 
been observed for any other SN~Ia at a similar phase.

Using the high quality observations and an empirical model to fit the optical filter 
light curves, we derive a continuous optical bolometric light curve for SN 1994D. With 
our continuous model of the filter light curves, bolometric magnitudes can be calculated
even when simultaneous measurements in all filter passbands are unavailable. We also
estimate the rise time to bolometric maximum and compare this value with predictions
from various models.

\section{The Empirical Model}

The shapes of the observed SN~Ia light curves (Leibundgut et al. 1991a; Riess et al. 
1996) suggest a simple analytical form. We model the time evolution of the observed 
brightness (in magnitudes in a given filter band) as a Gaussian (for the peak phase) 
atop a linear decay (for the late-time decline, $\sim$ 50 days past maximum light). 
A second Gaussian is introduced to fit the secondary maximum which is observed in the 
{\it V}, {\it R}, and {\it I} light curves about 25 days after the initial peak 
(Ford et al. 1993; Suntzeff 1996). To account for the rising (i.e., pre-maximum) 
segment of the light curves we multiply this model by a function which rises 
exponentially and approaches unity near the peak of the first Gaussian. The model 
parameters for the light curves in each of the available passbands are determined 
by a least-squares fitting technique. This simple approximation, with between 
5 and 10 free parameters, yields a surprisingly accurate representation of the 
optical filter light curves observed for most SNe~Ia (Vacca \& Leibundgut 1996a).
The model has the advantage of being completely independent of any assumptions 
regarding the nature of Type Ia events or possible similarities or differences 
from one event to another. Further discussion of the empirical model and the 
fitting procedures will be presented in a forthcoming paper (Vacca \& Leibundgut 1996b). 

We have used this empirical model to fit the light curve data for SN 1994D. 
In Figure 1, we show the model fits to the {\it U}, {\it B}, {\it V}, {\it R}, and 
{\it I} light curves for this supernova. 
The fits were restricted to within $\sim 100$ days of the maximum because many SNe 
have exhibited changes in the late-time decline several hundred days after maximum 
(Doggett \& Branch 1985; Leibundgut et al. 1991a). Errors on the data points in the 
{\it BVRI} passbands were assumed to be those given by Richmond et al.; errors on the 
$U$ band data were taken from Patat et al. (1995). The residuals from the fits are 
shown below each panel. Although our model fits the observed light curve data far 
better than the standard Type Ia templates of Leibundgut (1988), close inspection of 
the residuals reveals additional, small amplitude structure in the observed light 
curves. A flux excess relative to our model is present about 5 days after maximum 
in the {\it V}, {\it R}, and {\it I} light curves; this feature, which has 
never been reported before, is also seen in the unpublished data from CTIO and the 
CfA (H\"oflich 1995).

One of the most useful aspects of a continuous model of the light curve 
shapes is that additional interesting quantities can be 
objectively derived from it. In particular, we determine the rise time, $\Delta t_{30}$ 
(see below), the time of maximum brightness, $T_{max}$, the peak apparent magnitude, 
${\rm m}_{max}$, the magnitude difference between the peak and 15 days after the peak, 
$\Delta {\rm m}_{15}$, and the slope of the late-time decline, $s_2$.
The values of these parameters derived from our fits to the 
data for SN 1994D in each photometric band are given in Table 1.\footnote[3]{The results
presented in Table 1 supersede the values given in Vacca \& Leibundgut (1996a).}
Except for $s_2$, whose error was calculated as part of the least-squares
fitting routine, the uncertainties on all parameters derived from the best fit model 
(e.g, $T_{max}$ and ${\rm m}_{max}$) were estimated using
a Monte Carlo procedure. We simulated data points drawn from the best fit model 
at the same phases as the observed data points and with uncertainties estimated from 
the rms deviation of the observed data from the best fits.
The errors given in Table 1 correspond to the standard deviations measured 
after 1500 simulations.

\subsection{The Optical Bolometric Light Curve}

Using the results from the fits to the light curve data in the {\it UBVRI}
passbands, we constructed an optical bolometric, or ``quasi-bolometric'', light curve 
for SN 1994D. For this calculation we used the normalized passband transmission curves 
given by Bessell (1990) and zero points derived from the Kurucz (1992; see also
Castelli \& Kurucz 1994) model spectrum of Vega scaled to match the observed flux of 
Vega in the {\it V} band given by Hayes (1985). 
The integrated flux in each band was approximated by the mean flux multiplied 
by the effective width of the passband.
We corrected for the partial overlap of the {\it V}, {\it R}, and {\it I} bands,
and interpolated between the fluxes in the {\it U}, {\it B}, and {\it V} bands
in order to fill the small wavelength gaps between these passbands. We corrected
the fluxes for extinction by adopting the extinction values advocated by Richmond et al. 
(1995). The resulting quasi-bolometric light curve, the construction of which
is greatly facilitated by our continuous model of the individual filter
light curves, is shown at the bottom right of Figure 1. This curve represents the 
temporal evolution of the flux between $\sim 3280$ \AA\ and $\sim 8810$ \AA .
The secondary maximum observed in the {\it V}, {\it R}, and {\it I} data
carries through to the optical bolometric light curve. A similar feature is seen 
in the bolometric light curve of SN 1992A and possibly in SN~1991T (Suntzeff 1996). 
In fact, the shape of the bolometric and filter light curves of SN 1994D are very 
similar to those of SN 1992A.
We fitted the optical bolometric light curve with our model and determined 
its parameters and their uncertainties, which are listed
in Table 1. The empirical model reproduces the quasi-bolometric
light curve to better than $0.01$ mag over the entire temporal range.

The maximum flux for the quasi-bolometric light curve was found to be 
$3.9 (\pm 0.1) \times 10^{-10}$ erg/cm$^2$/s. For a distance of $13.7$ Mpc 
(Richmond et al. 1995; Patat et al. 1995), this implies a maximum luminosity of 
$8.8 (\pm 0.2) \times 10^{42}$ erg/s. As shown by Suntzeff (1996), the flux in the 
{\it U} through {\it I} photometric bands comprises $\sim 80$\% of the thermal 
energy radiated by a SN Ia between $2000$ \AA\ and $2.2 \mu$, and very little additional
flux is expected to be emitted at wavelengths $\lambda < 2000$ \AA\ and 
$\lambda > 2.2 \mu$ (Blair \& Panagia 1987).\footnote[4]{This provides
some justification for our use of the term ``quasi-bolometric''.} From the results given
by Suntzeff (1996) for SN 1992A, the flux in the near-infrared wavelength bands is expected 
to contribute less than $\sim 15$ \% to the total flux even at $80$ days after maximum. 
>From the {\it J}, {\it H}, and {\it K} magnitudes given by Richmond et al. (1995) 
corresponding to about six days after the maximum of the quasi-bolometric curve, we 
estimate that the flux in these near-infrared bands represents less than $3$ \% of 
the total flux in the combined optical bands. Based on assumed intrinsic colors, 
Leibundgut (1996) estimated that the infrared contributes less than 5\% near maximum light.
An upper limit to the true bolometric correction can be estimated from
the total flux for a blackbody with optical colors equal to those observed for SN 1994D. 
We fitted the observed fluxes in the {\it UBVRI} passbands with a Planck spectrum 
and derived a bolometric correction from the best fit temperature. Although a blackbody 
spectrum is generally a poor fit to the observed spectral energy distributions 
we use it only to obtain a rough estimate of the bolometric correction. 
At the optical bolometric maximum, the blackbody effective temperature was 
found to be $\sim 14000$ K and the bolometric correction, in terms of the ratio of total
flux to flux in the optical passbands, is a factor of $\sim 2.7$. 
With this value we can place limits on the peak bolometric
flux from SN~1994D of $42.9 < \log [L_{bol}\cdot (\frac{D}{13.7 Mpc})^2] < 43.3$. 
We note that during those phases when the Planck
function did provide a good fit to the observations (during the pre-maximum 
rise), the bolometric correction was found to be about a factor of $2$. It should 
also be noted that although we have not accounted for energy lost due to freely
escaping gamma rays in our calculations, we expect this energy loss to
represent only a very small fraction of the bolometric luminosity at
maximum (Leibundgut \& Pinto 1992).

The luminosity range derived above extends up to $\log L_{bol} < 43.5$ and $43.8$
for Virgo distance moduli of 31.1 and 32.0, respectively. This range
encompasses all luminosity predictions for models of the expolsion of Chandrasekhar-mass 
white dwarfs (Woosley \& Weaver 1994, 1995; H\"oflich \& Khokhlov 1996; Eastman 1996). 
For the Virgo cluster distance modulus favored by Sandage et al. (1996), SN 1994D 
would be highly overluminous with respect to the current models for Type Ia explosions.

\subsection{Rise Times and Times of Maximum}

The rise time, defined here as the difference (in days) between the time of maximum 
brightness and the time when the supernova was 30 magnitudes below maximum, $\Delta t_{30}$, 
was found to be $17.8 \pm 1.7$ days for the optical bolometric light curve. Similar values
are found for the rise times of the individual filter light curves (Table 1). The magnitude 
difference was chosen to correspond roughly to the difference in luminosity from the 
pre-explosion white dwarf to the peak of a typical SN~Ia. Because the light curve is so 
steep before maximum, the estimated rise time is relatively insensitive to the precise value 
of the magnitude difference chosen. It is critically dependent on the assumed shape 
of the light curve prior to maximum, however. We have assumed that the light curve rises 
(that is, the apparent magnitudes decrease) exponentially with time. It is certainly
possible that the pre-maximum light curves of SN Ia's rise faster than this, in which
case our procedure will yield an overestimate of the rise time. Additional pre-maximum data
for other SNe, as well as comparisons with detailed model calculations, will be needed to 
determine if our assumed form is an accurate representation of the pre-maximum segment of 
SN Ia light curves.

We note that our estimate of the rise time of the {\it B} light curve for SN 1994D 
is very similar to the value determined for SN 1990N ($17 - 20$ days) by Leibundgut 
et al. (1991b) using a different method. (A bolometric light curve 
constructed for SN 1990N that includes satellite ultraviolet data was found to have 
a bolometric rise time of about $14-18$ days [Leibundgut 1996].) We applied our fitting
procedure to the {\it B} data for SN 1990N and derived a rise time of $\sim 19$ days, 
a result which indicates that we are not grossly overestimating the rise time with 
our procedure and provides some confidence in our values for SN 1994D. Furthermore, based on 
the estimated time of maximum and the first observation of SN 1994D in the {\it R} band, 
the rise time in {\it R} must be longer than $\sim 13.3$ days. We find a similarly long
rise time when we apply our procedure to the {\it B} data for SN 1991T ($\sim 18.5$ days; 
Vacca \& Leibundgut 1996b). The rise times of the {\it V} light curves for 
SN 1994D and SN 1992A are also found to be approximately the same ($\sim 20$ days). 

For SN 1994D, the estimated rise time corresponds to an explosion date of $\sim$ 1994 
March 3 (JD 2449414.6), consistent with the values obtained from the individual filter 
curves. The first observation on 7 March was therefore only $4-5$ days after the explosion. 
Our estimated explosion date agrees surprisingly well with that found by H\"oflich 
(1995), although our estimate of the rise time to bolometric maximum is $\sim 3$ days longer. 

Contrary to the expectations from the models or the template light curves of 
Leibundgut (1988), we find only small differences between the times of the maxima 
in the {\it B}, {\it V}, {\it R}, and bolometric light curves. They all occur within 
about one day of one another (Table~1). In addition, the {\it U} and {\it B} maxima
are separated by only $\sim 1$ day. Similarly small differences in the times of 
maximum for the various filter light curves are found for several other well-observed 
SNe~Ia (e.g., SN 1989B and SN 1990N; see Vacca \& Leibundgut 1996a, 1996b).
Interestingly, the dim and fast-declining events SNe~1986G and SN~1991bg exhibit a 
delay of about $2$ days between the maxima in {\it B} and {\it V}.

\section{Discussion and Conclusions}

The extensive photometric observations of SN 1994D, along with a fairly simple
empirical fitting procedure applied to the individual filter light curves, have 
allowed us to construct a continuous ``quasi-bolometric'' light curve and to estimate 
the rise time for this SN Ia. The fits yield a far more objective means of quantifying 
the parameters describing the light curves than the standard templates can provide.
Because the procedure can be applied to the filter light curves of any individual SN Ia, 
it also can provide an independent test of the commonly adopted 
assumption that all SNe~Ia are members of a single family (Hamuy et al. 1995; 
Riess et al. 1996). While providing a close approximation to the light curve data, 
the fits can also reveal small amplitude ($< 0.05$ mag) substructure in the 
light curves which might escape notice if the light curve templates were adopted.

Our fit to the optical bolometric light curve of SN 1994D yields a peak luminosity in 
the wavelength range $3300$ to $8800$ \AA\ of $8.8 (\pm 0.2) \times 10^{42} \cdot 
(\frac{D}{13.7 Mpc})^2$ erg/s.
After accounting for possible bolometric corrections we find 
the peak bolometric luminosity to be in the range  
$42.9 < \log [L_{bol}\cdot (\frac{D}{13.7 Mpc})^2] < 43.3$. This corresponds to a mass of
$^{56}$Ni produced in the explosion of $0.4 < M_{Ni} < 1.1~M_\odot$ 
(Arnett, Branch, \& Wheeler 1985).  
The bolometric light curve also exhibits 
a very long rise time to maximum of $\sim 18$ days and an inflection or ``shoulder'' 
about 25 days after the primary peak ($\sim 40$ days after the explosion). 

Although the estimated peak bolometric luminosity for SN 1994D is in good agreement
with the range of model predictions, it is not clear if any of the current explosion models 
can reproduce the estimated rise times or the detailed structure of the observed filter or 
bolometric light curves for SN 1994D. 
Most current explosion models have difficulty in reproducing the long rise times found 
for SN 1994D and SN 1990N. Direct detonation and deflagration models 
yield rise times which are on the order of $7 - 14$ days (e.g., Khokhlov et
al. 1993). Delayed detonations may have rise times as long as $\sim 15$ days, but only
pulsating delayed detonation models or
detonation models incorporating a low density damping envelope have rise times 
as long as that determined for SN 1994D (H\"oflich \& Khokhlov 1996). 
None of the delayed detonation models considered by H\"oflich (1995)
or the pulsating delayed detonation models investigated by H\"oflich, Khokhlov, 
\& Wheeler (1995) can reproduce both the slow rise time and the peak 
bolometric luminosity of SN 1994D (unless it was substantially closer than $13.7$ Mpc).
The opacity and the treatment of the radiative transfer problem
during the initial stages of the explosion have a critical influence on the rise times 
in the various models. However, because of poorly known oscillator strengths and incomplete
line lists, the opacity is a source of uncertainty in the model calculations
(Khokhlov et al. 1993; Eastman 1996). Unless damping envelopes are fairly common,
the long rise times found for the light curves of SN 1994D and other SNe Ia may 
indicate that the opacity has been greatly underestimated.

The inflection seen in the bolometric light curves of SN~1994D and SN 1992A about 25 days 
after maximum is reproduced in some of the pulsating delayed detonation models for the 
explosion of Chandrasekhar-mass white dwarfs (H\"oflich, Khokhlov, \& Wheeler 1995; 
Eastman 1996). The presence of this ``shoulder'' in the predicted light curves 
from the explosions of sub-Chandrasekhar-mass white dwarfs depends on the
details of the model calculations (Eastman 1996; H\"oflich \& Khokhlov 1996).

Although beyond the scope of this paper, a detailed comparison between the bolometric 
light curve we have constructed for SN 1994D and those predicted by theoretical models 
would clearly represent an important test of various Type Ia explosion models. 
As the data for SNe Ia continue to improve in quality and quantity, the features observed in 
the light curves of SNe Ia's, and their variation from one supernova to another, will ultimately 
provide a means of distinguishing the correct models for these events.
Our light curve fitting method, by providing a quantitative description of the 
characteristics of individual observed light curves, should make the comparison between 
observations and models very straightforward.

\acknowledgments
We thank Mark Phillips and Nick Suntzeff for providing their light curve data for SN 1992A. 
WDV acknowledges support from the Beatrice Watson Parrent Foundation.     

\clearpage

\begin{deluxetable}{l r r c c c}
\scriptsize
\tablecaption{Derived Parameters for SN 1994D Light Curves}
\tablehead{\colhead{Band}
& \colhead{$\Delta t_{30}$} & \colhead{$T_{max}$} & 
  \colhead{${\rm m}_{max}$} & $\Delta {\rm m}_{15}$ & $s_2$ \\
     & \colhead{(days)} & \colhead{(day)\tablenotemark{a}} & 
  \colhead{(mag)} & \colhead{(mag)} & \colhead{(mag/day)}}
\startdata
{\it U} &  $16.1 \pm 0.7$ & $431.70 \pm 0.14$ & $11.31 \pm 0.02$ & 
$1.82 \pm 0.03$ & $0.0166 \pm 0.0032$ \\
{\it B} &  $17.6 \pm 0.5$ & $432.90 \pm 0.15$ & $11.85 \pm 0.01$ &  
$1.47 \pm 0.03$ & $0.0166 \pm 0.0002$ \\
{\it V} &  $20.1 \pm 0.6$ & $433.41 \pm 0.15$ & $11.90 \pm 0.01$ &  
$0.83 \pm 0.02$ & $0.0311 \pm 0.0003$ \\
{\it R} &  $19.3 \pm 0.6$ & $432.21 \pm 0.21$ & $11.87 \pm 0.02$ &  
$0.81 \pm 0.03$ & $0.0342 \pm 0.0002$ \\
{\it I} &  $20.0 \pm 3.0$ & $429.79 \pm 0.25$ & $12.10 \pm 0.02$ &  
$0.64 \pm 0.04$ & $0.0523 \pm 0.0001$ \\
Bol.    &  $17.8 \pm 1.7$ & $432.35 \pm 0.27$ & $12.04 \pm 0.02$ &  
$1.17 \pm 0.03$ & $0.0305 \pm 0.0008$ \\
\enddata
\tablenotetext{a}{JD $- 2449000$}
\end{deluxetable}

\clearpage

\clearpage
 
\begin{figure}
\plotfiddle{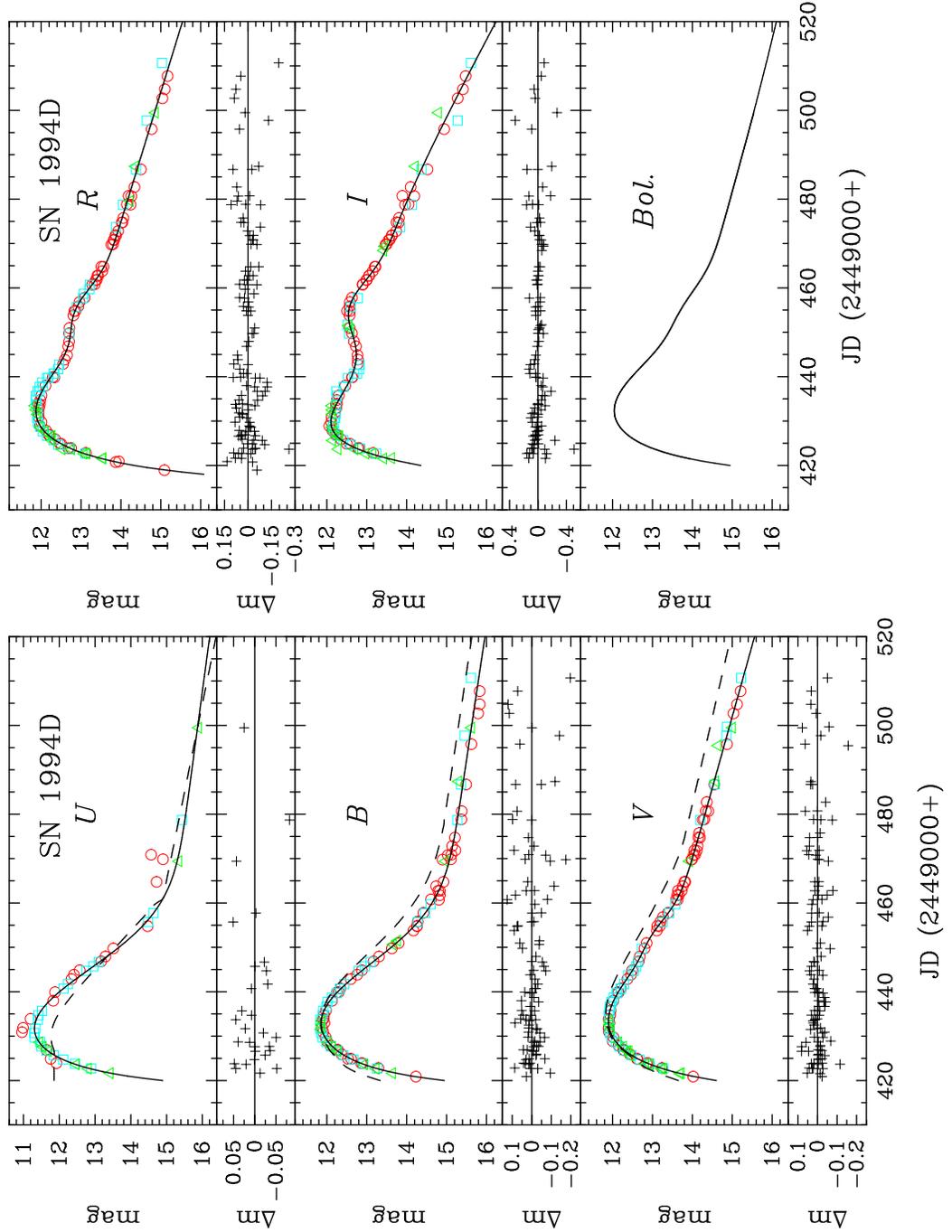}{18cm}{0}{70}{70}{-190}{-10}
\figcaption{$U$, $B$, $V$, $R$, $I$, and bolometric light curves of SN 1994D. Symbols
are the observed data from Richmond et al. 1995 ({\it circles}), Patat et al. 1995 
({\it squares}), and Meikle et al. 1996 ({\it triangles}). The solid lines are the best 
fitting models to the data points in each passband. Residuals are shown below each
panel. The dashed lines are the light curve templates of Leibundgut (1988) normalized 
to the time and magnitude of the maximum light in $B$.}
\end{figure} 

\clearpage

\clearpage
\end{document}